# A Modular IoT-Enabled Remote Laboratory Platform for Hybrid Energy System Research and Engineering Education

**Lamine Chalal* Louis Olivier* Pierre Liennard***
**Allal Saadane* Ahmed Rachid****

* *Icam School of Engineering, Lille Campus, 6 Rue Auber, B.P 10079, CEDEX, 59016 Lille, France (e-mail: lamine.chalal@icam.fr, louis.olivier@2026.icam.fr, pierre.liennard@2026.icam.fr, allal.saadane@icam.fr)*
** *Laboratory of Innovative Technologies, University of Picardie Jules Verne, 80000 Amiens, France (e-mail: rachid@u-picardie.fr)*

**Abstract:** Remote laboratory systems improve accessibility in engineering education and research by enabling Internet-based interaction with physical equipment. This paper presents a modular IoT-enabled remote laboratory platform for hybrid energy system studies, combining renewable energy emulators, battery storage, and programmable loads within a three-interface architecture based on a web HMI, TIA Portal, and MATLAB/Simulink, all connected through a Talk2M VPN cloud. An industrial PLC and IoT gateway provide deterministic local control as well as secure remote access and monitoring. A hierarchical energy-management algorithm is validated by comparing local and remote executions under identical wind and irradiance profiles. The results show small differences in the energy balances of the renewable sources, battery, and load, while typical communication delays are on the order of 100 ms. Consequently, the platform supports research-grade remote experimentation and project-based learning in control and energy systems engineering.

*Keywords:* Remote control, Remote laboratory, Hybrid energy systems, Industrial IoT, PLC-based control, Digital twin, Control education laboratories

## 1. INTRODUCTION

The global energy transition demands engineers capable of designing, implementing, and managing complex hybrid energy systems. As power systems increasingly integrate distributed renewable resources such as photovoltaic and wind energy, together with storage and flexible demand, new competencies are required to manage these systems efficiently and securely Kabeyi and Olanrewaju (2023).

Conventional laboratories provide valuable hands-on learning, but their accessibility is limited by equipment cost, safety constraints, scheduling, and maintenance requirements, especially across geographically distributed institutions Achuthan et al. (2021); Santana et al. (2013). These constraints are particularly critical when working with high-power or high-voltage equipment, where safety and supervision requirements further limit the number and duration of on-site sessions.

Over the last two decades, remote laboratories (RLs) and digital twins (DTs) have emerged as complementary tools to mitigate these limitations by enabling safe, repeatable, and geographically independent interaction with physical equipment for both education and research Mohammed et al. (2020); Ticona-Zela et al. (2022); Abekiri et al. (2023); Lei et al. (2024). As reported in Cappelaere and Saadane (2004), early RL deployments mostly targeted small-scale educational scenarios. Their scope has progressively expanded towards more complex cyber-physical systems and energy applications. The COVID-19 pandemic acted as a strong accelerator, forcing many institutions to rapidly deploy remote access to laboratory resources in order to maintain hands-on activities during campus closures.

According to the Scopus-based publication trend shown in Fig. 1, the field of RLs has experienced four distinct phases: initial growth (2000–2012), stagnation (2012–2016), rapid expansion (2017–2021) peaking during the COVID-19 period, and recent normalization (2022–2024). Despite this widespread use in education, RLs remain less developed for collaborative research applications Santana et al. (2013); Lei et al. (2024).

This paper addresses this gap by introducing a modular IoT-enabled RL platform designed specifically for hybrid energy systems. It combines industrial-grade control infrastructure with cloud connectivity, supporting both educational and research activities. The platform demonstrates how remote experimentation enables equitable access to advanced laboratory infrastructure, eliminating geographic barriers to participation. By providing secure, near real-time Internet-based access to the Icam demon-strator, the platform enables distributed researchers and students across partner institutions to conduct live experiments, validate control algorithms, and develop practical skills without requiring on-site presence or travel.

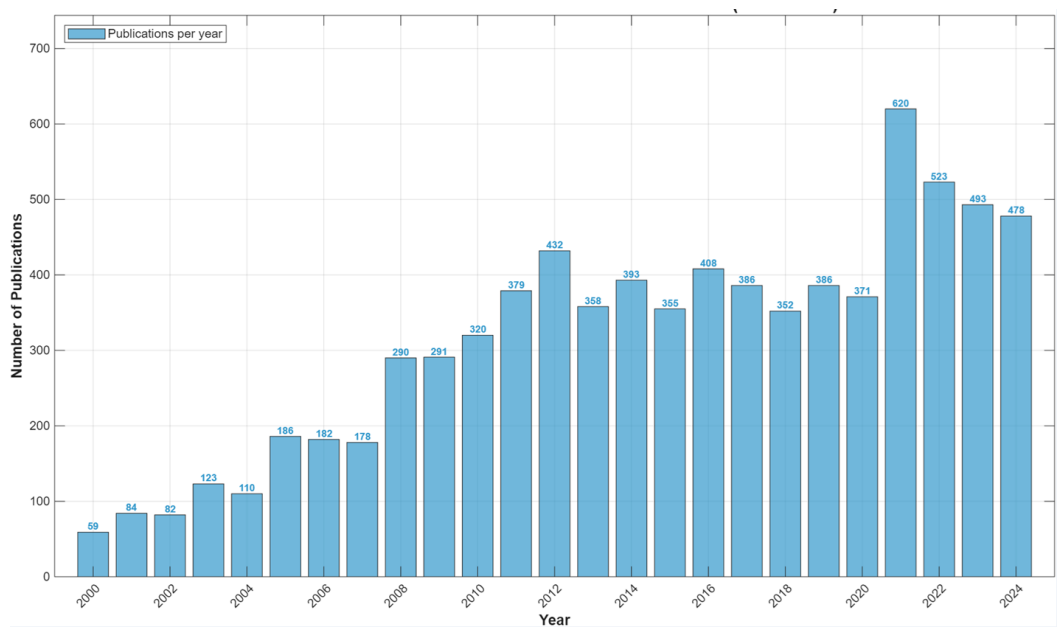

Fig. 1. Scopus-indexed publications on RLs (2000–2024).

Unlike earlier educational platforms such as EduSolar, which focus on low-power photovoltaic/thermal experiments Korkut and Rachid (2024), the proposed system targets research-grade hybrid microgrid operation with industrial PLC-based control and multi-interface access.

Specifically, the platform incorporates renewable energy emulators, energy storage systems, and adjustable loads, interconnected via power electronics interfaces (inverters, DC-DC converters), programmable logic controllers (PLCs), and IoT sensors, thereby enabling remote control and near real-time supervision of the entire demonstrator.

The main contributions are: (i) a modular RL architecture combining PLCs, IoT sensors, and multi-interface remote access; (ii) an experimental validation of reliability and local-versus-remote consistency for a hierarchical energy-management controller representative of typical microgrid strategies; and (iii) a platform supporting both hands-on learning and collaborative research, while extending a previously established digital-twin framework toward distributed experimentation.

The remainder of this paper is organized as follows: Section 2 reviews the evolution and state of the art in RL systems, identifying gaps in research applications. Section 3 describes the platform architecture, hardware components, and multi-interface control framework. Section 4 presents experimental validation demonstrating functional equivalence and reliability between local and remote operation. Section 5 concludes with key contributions, limitations, and opportunities for distributed collaborative research networks.

## 2. RELATED WORK AND LITERATURE REVIEW

RL systems have evolved from simple educational setups to complex infrastructures supporting both teaching and research. Early systems such as MIT iLabs Hardison et al. (2008), WebLab-DEUSTO Orduña et al. (2018), Library of Labs Richter et al. (2011), and LabShare Lowe et al. (2009) established the foundations of shared RL infrastructure in engineering education, reducing infrastructure costs through multi-institutional access while emphasizing pedagogical repeatability over research-grade configurability.

Most RL architectures combine user interfaces, hardware control layers, and communication software to provide Internet-based interaction with physical equipment and authentic experimental conditions for engineering education Pang et al. (2022); Achuthan et al. (2021); Kaluz et al. (2015).

More recent IoT-enabled and cloud-connected laboratories based on low-cost microcontrollers (Arduino, ESP32) and lightweight protocols (MQTT, WebSocket, Firebase) improve accessibility. Architectures such as LABER-SIME Abekiri et al. (2023) and the EduSolar photovoltaic/thermal lab Korkut and Rachid (2024) offer low-cost deployment, MQTT-based remote monitoring, and web-based dashboards, but remain primarily designed for low-power, teaching-oriented experiments rather than research-grade hybrid microgrid operation.

In parallel, a smaller number of research-oriented platforms have addressed advanced applications such as cyber-physical systems and hybrid energy systems. These infrastructures typically rely on MATLAB/Simulink, SCADA, or PLC controllers, but often suffer from interoperability issues, limited remote scalability, or insufficient multi-interface accessibility for distributed research networks Santana et al. (2013); Kaluz et al. (2015); Lei et al. (2024).

Table 1 synthesizes representative RL platforms, comparing their main characteristics with the system proposed in this work.

## 3. PLATFORM OVERVIEW AND HARDWARE CONFIGURATION

The experimental platform integrates a complete laboratory-scale hybrid energy system designed for research and educational applications in renewable energy management. The system adopts a modular architecture based on a typical multisource topology, wherein each functional unit operates as an independent, plug-and-play component. This modularity is realized through a relay matrix that enables dynamic reconfiguration of operational modes, allowing rapid topology switching and seamless hardware integration or upgrades. Table 2 summarizes the main hardware components and their specifications.

The platform combines real industrial devices (PLC, HMI, IoT gateway, smart meters, relay matrix, communication infrastructure) with programmable emulators for PV generation, wind input, and battery behavior. This choice ensures safe operation, repeatable test conditions, and independence from weather variability, while preserving realistic control, communication, and hardware-in-the-loop interactions. The modular architecture also allows future integration of additional real devices in parallel, depending on experimental objectives and power constraints.

The system operates in both on-grid and off-grid modes and supports experiments on converter control, renewable source prioritization, and energy management strategies. Its main features are modularity, secure IoT-based instrumentation, and support for progressive access from basic supervision to advanced control development.

Table 1. Comparative analysis of representative RL platforms

| Platform | Focus Area | Hardware Stack | Communication | Key Characteristics |
|---|---|---|---|---|
| **MIT iLabs** Hardison et al. (2008) | Education | DAQ + SCPI instruments | SOAP/AJAX web services | Scalable web-services architecture; multi-institutional sharing |
| **WebLab-DEUSTO** Orduña et al. (2018) | Education / Automation | PLC + DAQ modules | WebSocket-based access | Interactive and batched modes; federation-capable architecture |
| **Library of Labs** Richter et al. (2011) | Network sharing | Multiple heterogeneous platforms | SCORM / HTTP | 200+ standardized experiments; easy SCORM packaging and deployment |
| **Pang et al.** (DT) Pang et al. (2022) | Robotics + Education | PLC + DT models | WebSocket + IoT interfaces | ADDIE-based pedagogy; reported 80% effectiveness; real-time DT synchronization |
| **LABERSIME** Abekiri et al. (2023) | IoT + Education | ESP32-based embedded nodes | HTTP / cloud IoT services (LMS + IDE) | COVID-19 emergency response; low-cost deployment; not intended for hard real-time industrial control |
| **Santana et al.** Santana et al. (2013) | Automatic control | Parallel robot + PLC controller | Industrial network protocols | Real-time 3-DOF control; research-grade performance; SLD architecture |
| **Lei et al.** (DT) Lei et al. (2024) | Research + Education | Web-based DT linked to PLC/plant | PROFINET / OPC / VPN | Real-time physical–virtual synchronization; scalable, open-access architecture |
| **Proposed system** | Education + Research | Siemens S7-1200 + IoT sensors + emulators | PROFINET / Modbus / OPC UA / VPN | Modular, industrial-grade hybrid energy platform; multi-interface HMI; builds on prior digital-twin integration |

Table 2. Core hardware components of the hybrid energy laboratory platform.

| Subsystem | Equipment / Models | Main specifications | Interface / Control |
|---|---|---|---|
| **PV generation** | PV emulator (EA–PSI 9360–15 2U); artificial light array; PV modules (SPR-215-WHT) | 0–360 V, 0–15 A programmable DC source; 4 kW dimmable lamps; 215 Wp per panel | PV emulator via Modbus TCP/IP; lamps via analog 0–10 V; PV modules on DC bus |
| **Wind generation** | Wind turbine emulator (EOLYS-500); permanent-magnet generator / rectifier | 400 W, 90 Vdc; 0–25 m/s equivalent wind speed; permanent-magnet generator feeding DC link | Wind emulator via analog 0–10 V; generator output connected to the DC bus through a rectifier |
| **Storage** | Battery emulator (EA–PSB 9200–70 3U) | 0–200 V, 70 A reversible source; programmable charge/discharge profiles | Ethernet / Modbus TCP/IP |
| **Loads** | Lighting loads (10 W, 70 W lamp banks); programmable resistive load | 0–300 W; selectable lamp groups; 0–100% resistive power range | Relay switching for lamps; analog 0–10 V control for dimmer |
| **Power electronics** | MPPT / inverter units (Victron, Steca) | 100/20 MPPT rating; DC/AC conversion for grid-connected or stand-alone operation | PWM / proprietary interfaces; coupled to DC bus and AC network |
| **Controller & HMI** | PLC Siemens S7-1200; HMI panel Siemens TP900 Comfort | PLCs with PROFINET and relay/0–10 V outputs; 9" touchscreen operator panel | PROFINET for PLC–HMI |
| **Sensing & remote access** | Smart meters (Lumel/Lumen P30H, P30P); weather station; Ewon Flexy 205 IoT gateway; IP cameras | AC/DC voltage, current, power and energy; wind, irradiance, temperature; VPN routing; video monitoring | Smart meters via Modbus (RTU/TCP); weather via local link + web API; gateway via Ethernet/VPN (Talk2M); cameras via HTTPS/RTSP |

*3.1 Multi-Interface Control Framework and User Access Levels*

The proposed control framework is structured around three complementary user access levels, as shown in Fig. 2. Each layer addresses specific educational or research needs while maintaining unified system control through the PLC and IoT gateway.

**Level 1 – Web Interface (Introductory Access):** ViewON, an Ewon software tool, is used to design and deploy web-based dashboards for remote access. These browser-based HMIs support real-time system monitoring, supervisory control, and remote interaction through a secure TLS-encrypted VPN connection. They also enable users to study the relationships between solar irradiance, temperature, and PV power output, as well as between wind speed and wind-turbine power generation, to perform energy balance assessments, and to analyze system efficiency, with real-time visual feedback provided via integrated IP cameras.

**Level 2 – TIA Portal Environment (Advanced Programming):** This level provides full access to IEC 61131-3 PLC programming and HMI configuration for advanced control and automation students. The S7-1200 PLC executes analog and digital I/O processing for emulator control, relay switching for hybrid system reconfiguration and renewable-source prioritization, as well as energy-management logic under variable meteorological conditions.

**Level 3 – MATLAB/Simulink Interface (Research Grade):** Bi-directional communication between MATLAB and the PLC is achieved via OPC UA, PROFINET, and Modbus for IoT sensors data, complemented by a web service API for weather data acquisition. This interface enables hardware-in-the-loop tests, algorithm deployment, and digital-twin validation Chalal et al. (2023), building on a previously published real-time synchronized digital-twin environment for hybrid energy systems and extending it here toward distributed remote experimentation. It also supports the development of advanced control algorithms (e.g., optimization, predictive control, machine-learning-based strategies) both locally and remotely.

This hierarchical design supports a progressive learning path from web-based supervision to advanced algorithmic experimentation and digital-twin validation, while preserving a single, unchanged physical setup.

### 3.2 Pedagogical and Research Applicability

The platform supports both semester-long student projects and research activities. Its hierarchical multi-interface architecture enables progressive skill development, ranging from web-based access for monitoring, analysis, and supervised experimentation to professional-grade industrial control environments.

The system has been deployed in semester-long projects involving multiple engineering students with exclusively remote access to the experimental infrastructure. Collected feedback indicates high satisfaction regarding remote access usability and perceived experimental realism, with reported improvements in understanding energy management, PLC programming, and industrial communication protocols. Current learning outcomes include distributed data acquisition, renewable energy systems integration, IEC 61131-3 standards compliance, and digital-twin development.

The Ewon Flexy gateway supports concurrent operation with up to five remote users, while a preconfigured remote desktop environment helps students without direct access to advanced control software (TIA Portal, MATLAB/Simulink). In addition, a detailed numerical model of the platform allows users to validate control algorithms offline and continue their work when the experimental setup is occupied or otherwise unavailable.

Beyond hardware safety, the platform includes several mechanisms to supervise remote sessions and protect the physical setup. Each remote experiment is associated with an explicit session duration, after which the system is automatically shut down unless the user actively extends the booking. In addition, a watchdog heartbeat signal exchanged via OPC UA between Simulink and the PLC allows the controller to detect loss of communication and safely disable the platform when supervision is no longer active. Layered protection mechanisms are also implemented at the PLC level, including thermal and operating limits for the PV and wind subsystems, as well as automatic shutdown of the artificial light source and wind emulator in the event of MATLAB–PLC or VPN connectivity loss. Taken together, these mechanisms ensure that remote access does not compromise the integrity of the physical platform.

The platform also enables distributed research collaboration across the international Icam network and partner institutions where local experimental facilities may be limited, by providing remote access to a shared industrial-grade demonstrator.

## 4. REMOTE EXPERIMENTATION ARCHITECTURE AND VALIDATION STRATEGY

### 4.1 Control Implementation: Hierarchical Framework

To assess system performance, a hierarchical control algorithm with multiple operating modes is implemented on the hybrid multi-source configuration integrating PV panels, a wind turbine emulator, battery energy storage, and DC/AC loads. The main objective is to ensure continuous load supply under varying meteorological conditions while maximizing renewable energy utilization. The corresponding mode-selection logic is summarized in Fig. 3.

**Energy Management Strategy:** The hybrid microgrid is governed by a priority-based controller that selects between PV-dominant, wind-dominant, and battery-support modes according to the available powers and battery state of charge. Renewable sources (PV or wind) are used preferentially, while the battery acts as an energy buffer, charging when surplus power is available and discharging when generation is insufficient, subject to SOC constraints. Supplying the local load remains the primary objective across all operating conditions.

### 4.2 Remote Experimentation and Validation

To validate remote operability, identical control experiments were executed locally and remotely by using the MATLAB/Simulink interface under the same wind speed and irradiance profiles (Fig. 4). Power profiles were recorded for the renewable sources, the battery, and the load, and the corresponding energy balances were obtained by integration.

Fig. 5 compares the resulting local and remote profiles for the renewable sources, the battery, and the load, together with the operating modes derived from the sequential control algorithm shown in Fig. 3. Across successive wind-, PV-, and battery-dominated phases, the controller maintained load supply and ensured smooth transitions between operating modes, while the remote trajectories (dashed lines) closely followed their local counterparts.

Table 3 presents the energy balance over the full test window for local and remote operation. The renewable sources delivered 18.07 Wh in local mode and 18.79 Wh in remote mode, while the battery exchanged energy was -16.66 Wh and -16.43 Wh, respectively. The load consumption reached 25.89 Wh locally and 26.55 Wh remotely. Deviations were 4.0% for the renewable sources, 1.4% for the battery, and 2.6% for the load. Negative battery values indicate a net discharging balance over the considered interval. The observed differences are mainly attributed to the communication latency introduced by the remote loop rather than to any modification of the control logic itself. For the class of experiments considered here, the remote implementation preserves the overall behavior of the energy-management algorithm with good fidelity.

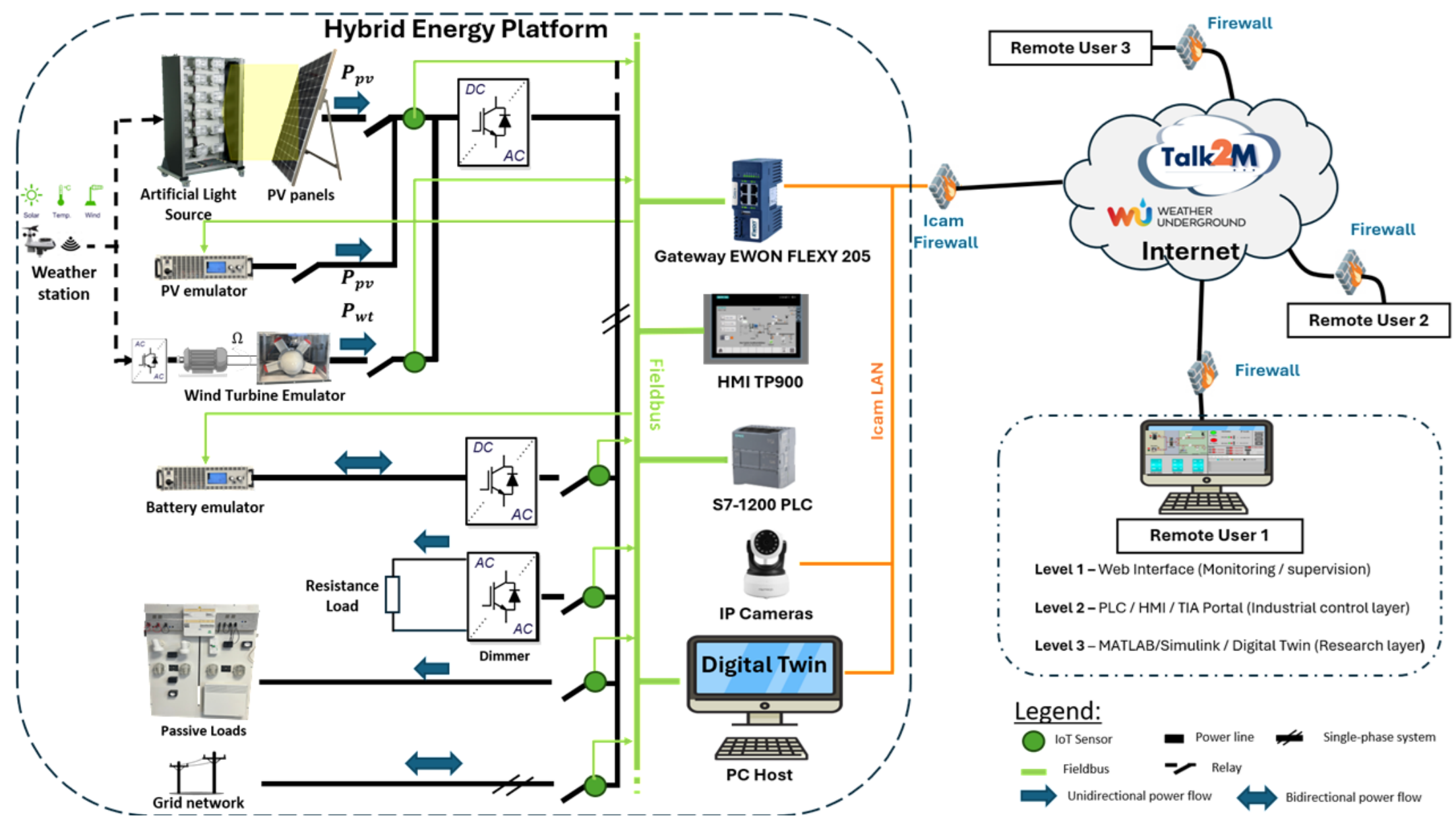

Fig. 2. Three-level multi-interface RL architecture, showing web-based access, PLC/TIA Portal supervision, and MATLAB/Simulink interaction through the IoT gateway and industrial control layer.

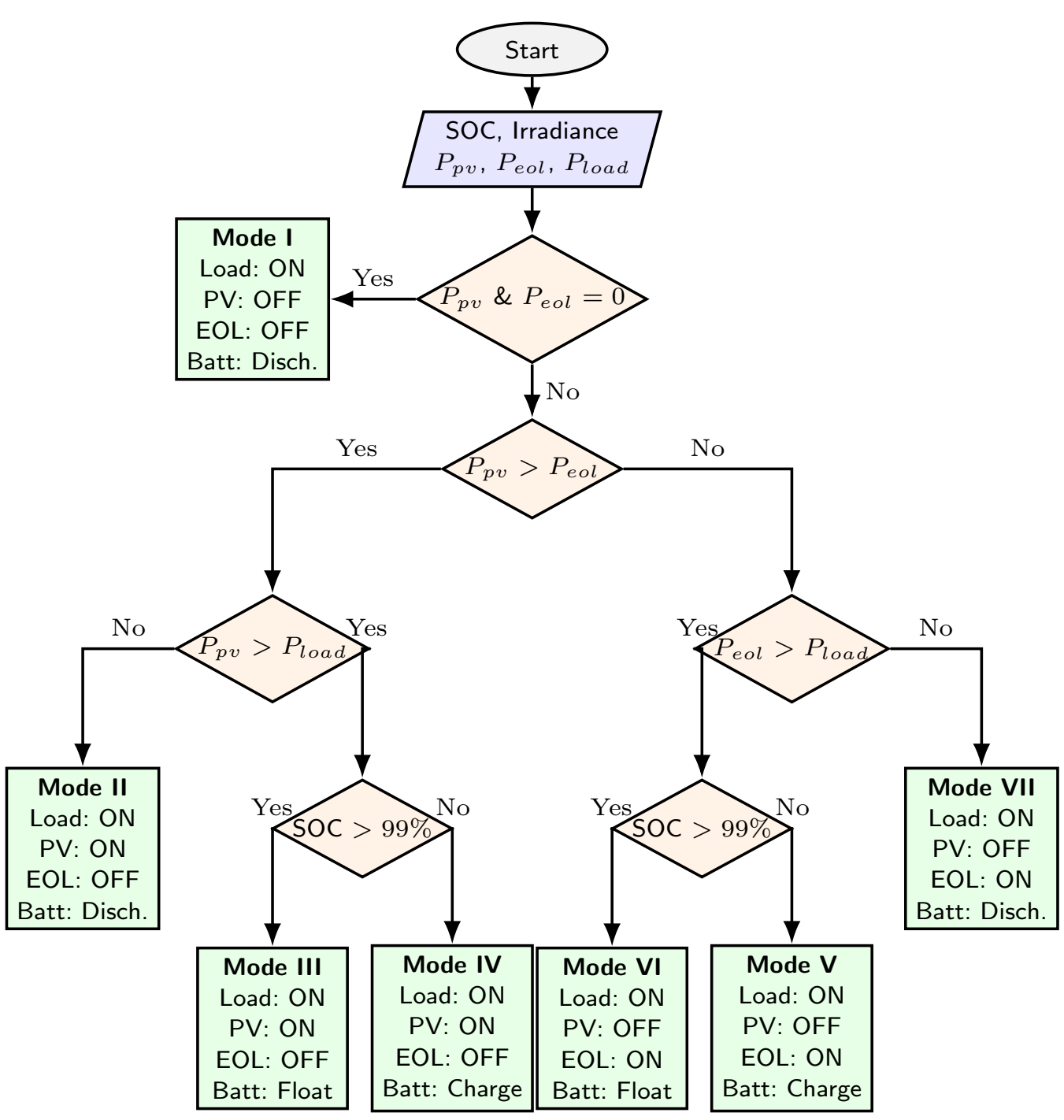

Fig. 3. Hybrid microgrid control mode selection algorithm.

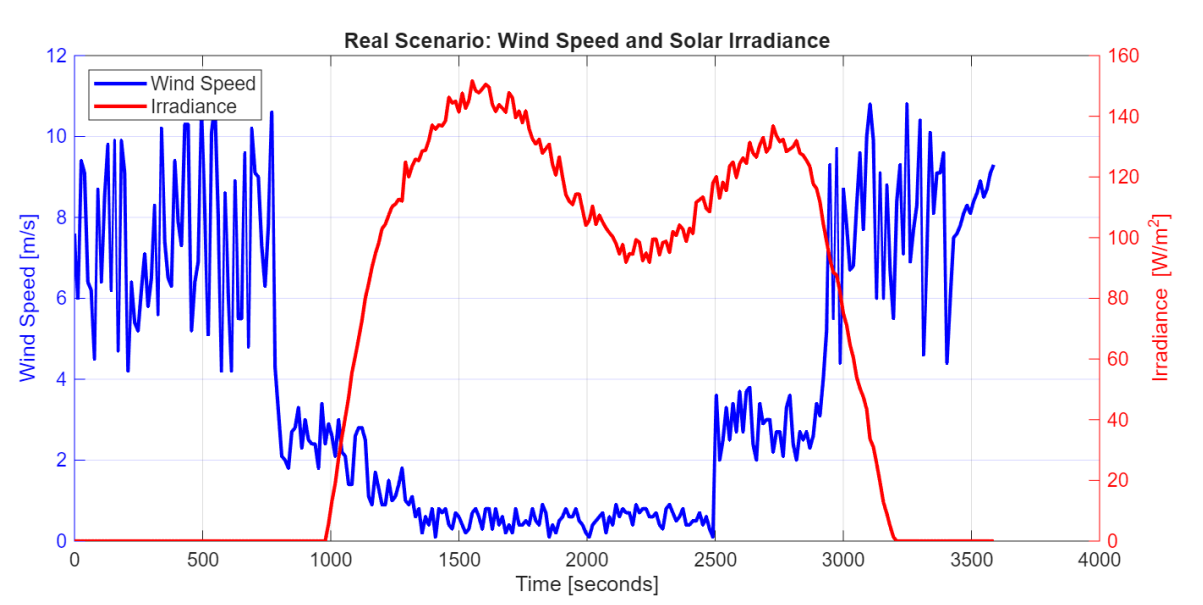

Fig. 4. Wind speed and solar irradiance profiles.

Table 3. Energy balance over the full test window.

| Term | Local (Wh) | Remote (Wh) | Dev. (%) |
|---|---|---|---|
| $E_{\mathrm{ren}}$ | 18.07 | 18.79 | 4.0 |
| $E_{\mathrm{bat}}$ | -16.66 | -16.43 | 1.4 |
| $E_{\mathrm{load}}$ | 25.89 | 26.55 | 2.6 |

Communication performance was evaluated over Internet links using the Talk2M VPN infrastructure. In remote mode, the exchanged data and control commands transit through the Icam LAN network, and the resulting communication delay may therefore be affected by local network traffic conditions. In local mode, control commands travel over the fieldbus network with very low latency. In remote mode, however, typical one-way communication delays are on the order of 100 ms. During the reported test campaign, the maximum observed latency peak was about 4 *s*.The cloud-based remote control loop should not be regarded as strictly real-time, although it performed adequately for the supervisory and validation experiments described here. In this context, the industrial VPN prioritizes encryption, tunneling, and connection robustness over latency minimization. Although these latency peaks did not compromise the experiments considered in this paper, they may become limiting for more time-critical closed-loop applications.

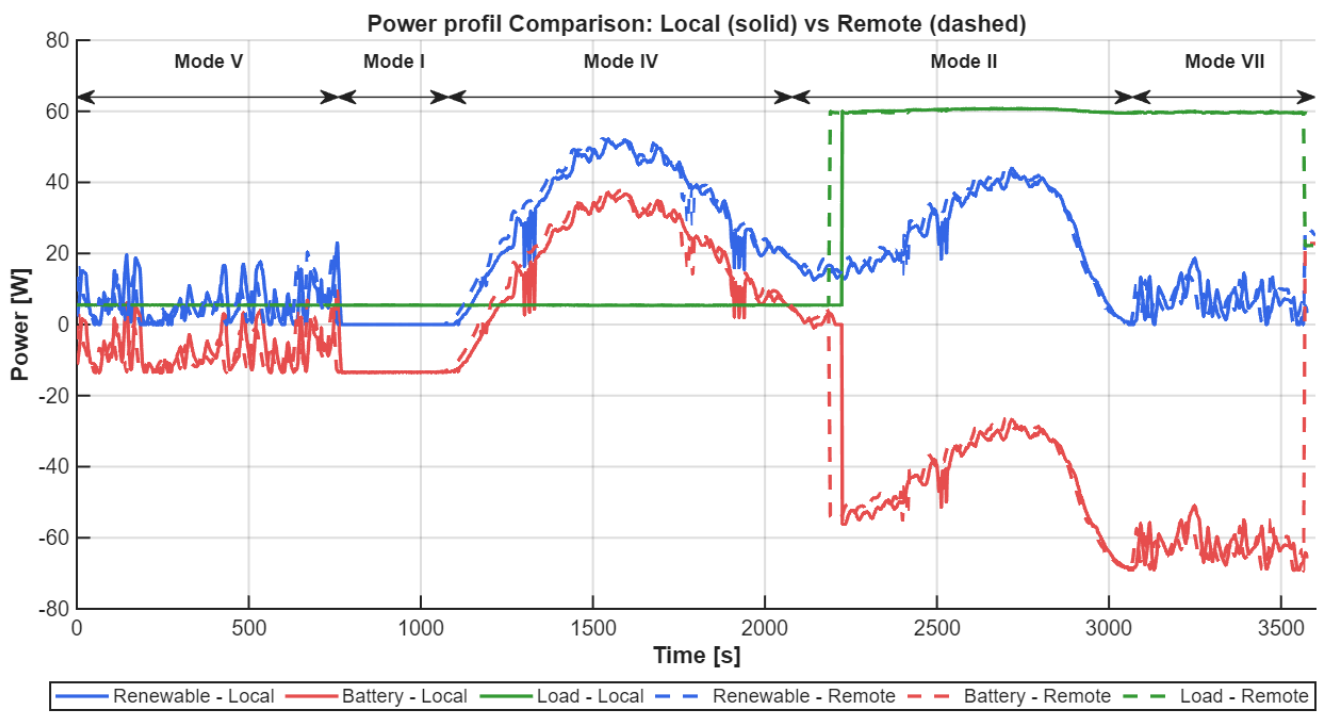

Fig. 5. Power flow comparison: local and remote operation.

## 5. CONCLUSION AND FUTURE WORK

This paper has presented a modular IoT-enabled remote laboratory platform for hybrid energy systems, supporting both engineering education and collaborative research. Comparative experiments under identical operating conditions showed only modest differences between local and remote execution, confirming the suitability of the platform for supervisory experimentation, controller validation, and project-based learning, while also highlighting that the current cloud-based access path is not intended for strictly real-time applications.

Future work will focus on lower-latency communication solutions, including direct 5G connectivity, more diverse experimental scenarios, and extensions of the digital-twin framework toward richer collaborative studies and advanced control strategies. The platform will also be extended toward electric-vehicle-oriented applications Chalal and Rachid (2025).

## DECLARATION OF GENERATIVE AI

During the preparation of this work, the authors used DeepL and ChatGPT to improve language and readability. The authors reviewed and edited the content and take full responsibility for the publication.